\documentclass[showpacs,preprintnumbers,amsmath,amssymb,twocolumn,pra]{revtex4-1}
\usepackage{amsfonts}
\usepackage{subfigure}
\usepackage{graphicx}
\usepackage{dcolumn}
\usepackage{bm}
\usepackage{times,epsfig,amssymb,amsmath}

\begin{document}

\title{Unbounded quantum Fisher information in two-path interferometry with finite photon number}
\author{Y. R. Zhang$^{1}$}
\email{yrzhang@iphy.ac.cn}
\author{G. R. Jin$^{2}$}
\author{J. P. Cao$^{1}$}
\author{W. M. Liu$^{1}$}
\author{H. Fan$^{1}$}
\email{hfan@iphy.ac.cn}
\affiliation{$^{1}$Institute of Physics, Chinese Academy of Sciences, Beijing 100190, China\\
$^{2}$Department of Physics, Beijing Jiaotong University, Beijing 100044, China}

\date{\today}
\pacs{42.50.St, 42.50.Dv, 42.50.Ex, 42.50.Lc}

\begin{abstract}
The minimum error of unbiased parameter estimation is quantified by
the quantum Fisher information in accordance to the Cram\'{e}r-Rao
bound. We indicate that only superposed NOON
states by simultaneous measurements can achieve the maximum quantum
Fisher information with form $\langle\hat{N}^{2}\rangle$ for a given
photon number distribution by a POVM in linear two-path interferometer
phase measurement. We present a series of specified superposed states
with infinite quantum Fisher information but with finite average
photon numbers. The advantage of this unbounded quantum Fisher information
will be beneficial to many applications in quantum technology.
\end{abstract}

\maketitle

\section{Introduction}\label{I}
\begin{figure}[!hbt]
\subfigure[]{
\includegraphics[width=0.35\textwidth]{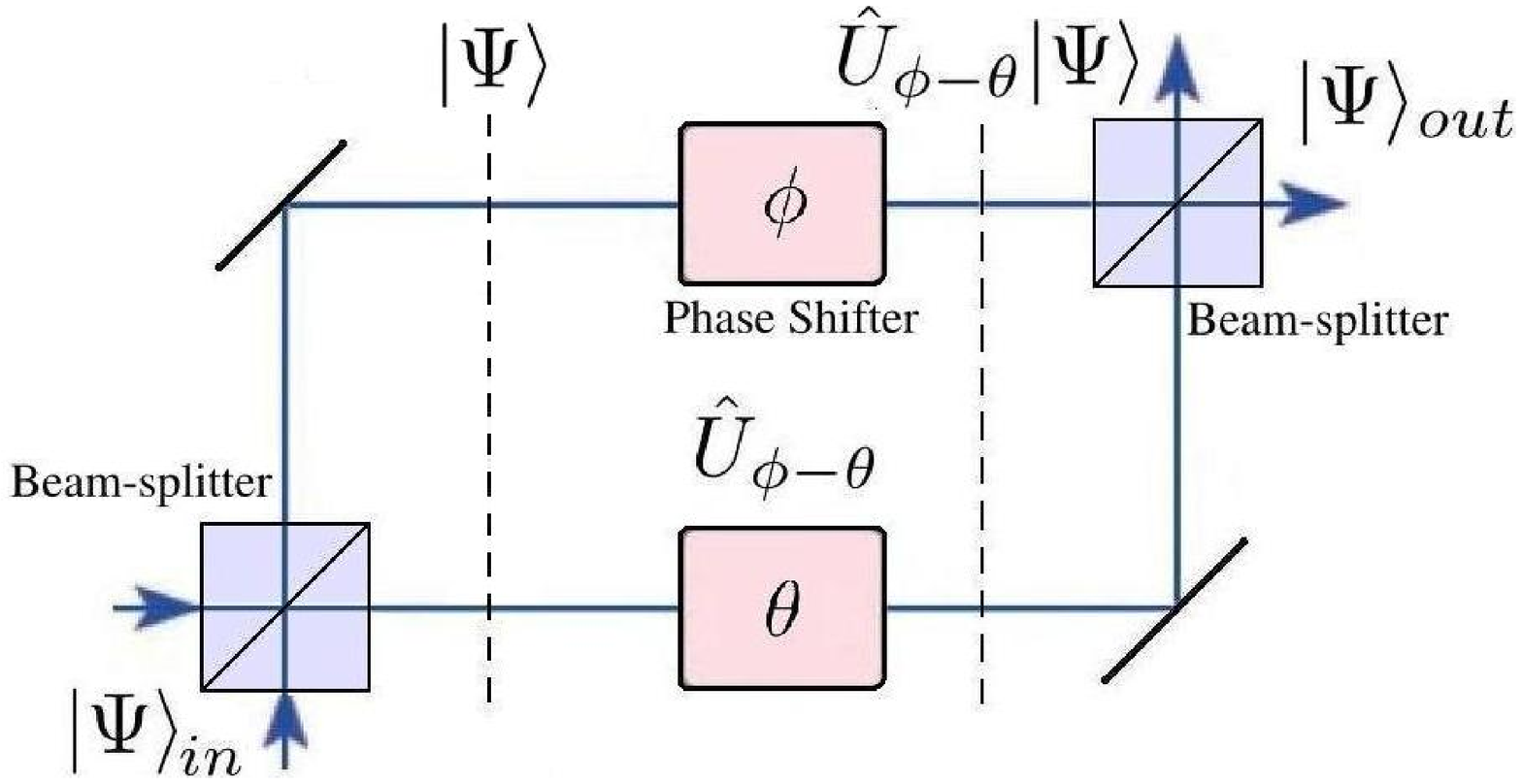}\label{f1a}}
\subfigure[]{
\includegraphics[width=0.35\textwidth]{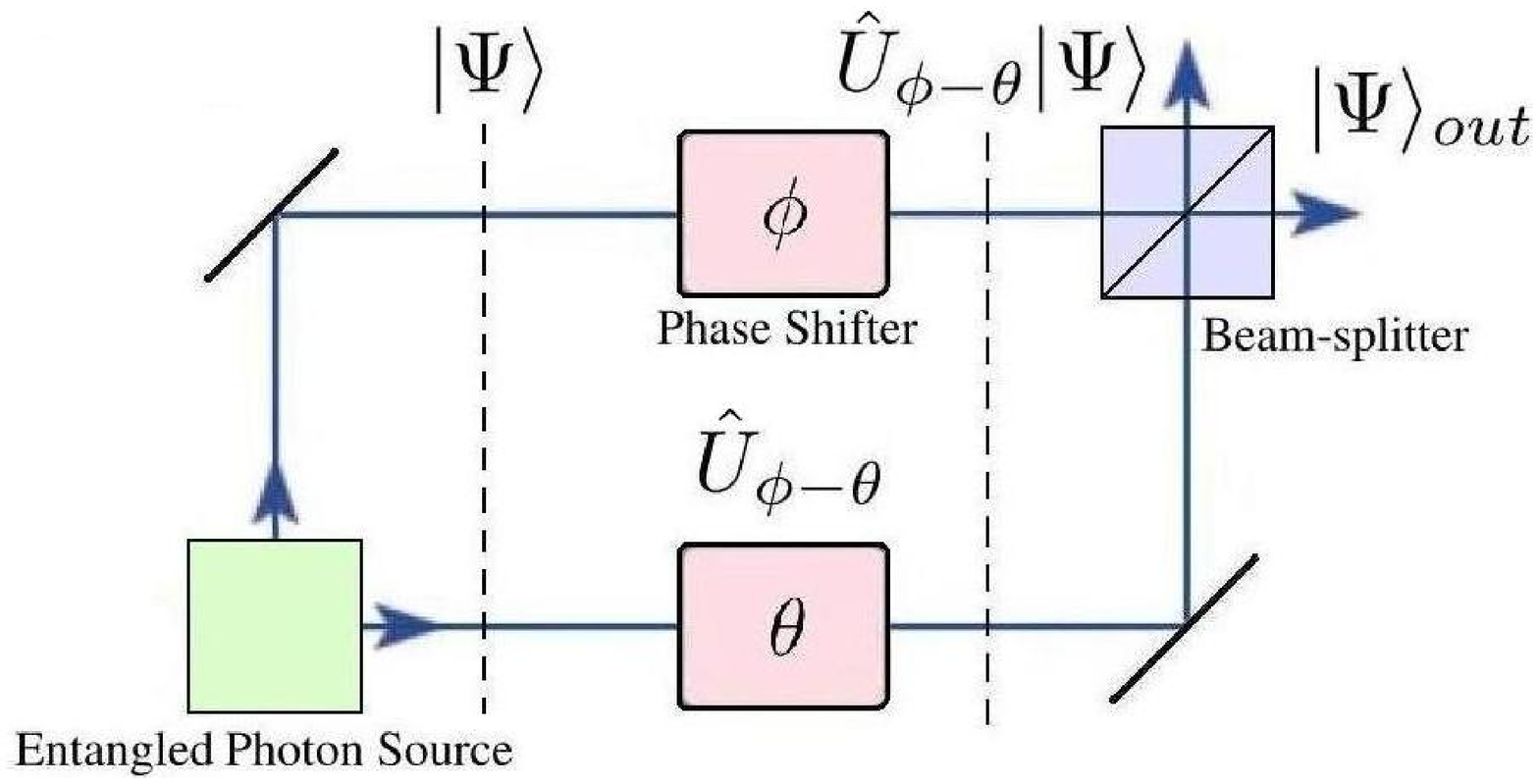}\label{f1b}}
\caption{\label{f1}(color online). (a)\ Setup of Mach-Zehnder Interferometer.
$|\Psi\rangle_{in}$ is the input state. Both beam-splitters are
$50:50$.
The unknown phase shift $\phi$ is to be
estimated. A controllable phase $\theta$, in the other arm, is used
in the experiment to achieve the best precision.
$\hat{U}_{\phi-\theta}$ is the phase operator and
$\hat{U}_{\phi-\theta}|\Psi\rangle$ is the intermediate state.
Passing through the second beam-splitter is the output state
$|\Psi\rangle_{out}$. We perform photon number counting measurements
at two ports of Mach-Zehnder Interferometer to measure the output
state, respectively. (b)\ Setup of modified Mach-Zehnder Interferometer.
The first beam-splitter is replaced by an entangled photon source.
We assume that the entangled photon source can produce NOON states and
superposed NOON states.}
\end{figure}

Precise interferometric measurement plays a key role in many scientific
and technological applications, such as quantum metrology, imaging, sensing
and information processing \cite{book}. The fundamental sensitivity bounds
of phase measurement in the Mach-Zehnder interferometer (MZI), as shown in
Fig.\ \ref{f1a}, are of broad interest for those areas. For $N$ identical
uncorrelated particles, the error about the phase measured in MZI on the
average decreases as $1/\langle\hat{N}\rangle^{1/2}$ which is the
shot-noise limit (SNL). In 1980s, it was pointed out that by using coherent
light together with squeezed vacuum we could beat SNL \cite{Caves}. It is
also shown that using NOON state \cite{NOON state} and quantum entanglement
allows interferometric sensitivity that also surpasses this limit. Instead
of SNL, the ultimate limit imposed by quantum mechanics is the Heisenberg
limit (HL) with a generally accepted form $1/\langle\hat{N}\rangle$
\cite{ouZY,LloydSci}. Experiments exploring those topics have been performed
in various systems with photons \cite{USTC,JpSci}, ions \cite{ions},
cold-atoms \cite{cold} and Bose-Einstein condensates \cite{BEC,BEC2}. However,
it seems to be indicated in some works that the HL with form
$1/\langle\hat{N}\rangle$ could be violated while the experiments are
performed with a fluctuating number of particles \cite{Dowling,cite}. Then,
HL for an unfixed number of particles has become a focus of great attention
as well as contention in all the parameter estimation schemes.

In this Letter, we investigate the quantum Fisher information (QFI) in two
kinds of two-path interferometers for a fluctuating number of photons since
the well-known Cram\'{e}r-Rao bound (CRB) whose leading role is played by
QFI is often used to estimate sensitivity of phase measurement. To achieve
the maximum QFI and the lowest CRB, we should use the optimal states, and
implement the optimal measurements (optimal POVM). We investigate the
measurement applied in two kinds of linear two-path interferometers, as
shown in Fig.\ \ref{f1}. Note that Fig.\ \ref{f1a} is for MZI, and
Fig.\ \ref{f1b} is a modified Mach-Zehnder interferometer (MMZI) where 
the first beam-splitter is replaced by an entangled photon source. We 
show that the optimal measurement scheme can be expressed by a group 
of compatible observables. We confirm that the maximum QFI for a 
definite photon number probability distribution, written
as $\langle\hat{N}^{2}\rangle$, is saturated with a superposition of
NOON states in MMZI. In particular, we present a superposition of NOON states
with a specified probability distribution and a finite average photon number.
We find that QFI for this state can be infinite, i.e. an arbitrary high phase
sensitivity is obtained by Cram\'{e}r-Rao inequality while it still has a
finite average photon number. Apparently, this phase sensitivity does not
violate Heisenberg uncertainty relation, but seems to break HL.

Nevertheless, CRB is only asymptotically tight for infinitely many trials
of measurements under the unbiased estimate assumption \cite{extra1,arxiv},
and it includes no prior information of the phase probability distribution
\cite{zzb}. Therefore, HL can merely be violated under some very special
situation, for example the phase is in a restricted neighborhood near zero
as the case of distinguishing states, and the limit obtained by CRB is of
little use when we consider
the resources required for the prior information of the problem
\cite{subHL,OHL,NJP}. Similar theoretical results based on the quantum speed
limit \cite{MLtheorem} are also obtained in Ref.\ \cite{newHL}. Although
CRB can sometimes grossly underestimate the achievable error, QFI and CRB
are still of great use for many applications in quantum techniques such as
parameter estimation for noisy systems \cite{noise1,noise2}, limits of
imaging \cite{QIma}, distinguishability of states \cite{Caves2} and
quantum Zeno effect \cite{QZE}. Related topics and recent developments
can be found in
Refs.\ \cite{OpC,new,triphoton,Nolinear,NC,chipPRL,Hofmann2,extra2,extra3,extra4}.

This paper is organized as follows. In Sec.\ \ref{II}, we demonstrate
the optimal measurement scheme by simultaneous measurements in linear
two-path interferometers. In Sec.\ \ref{III}, we obtain the maximum QFI
in the optimal measurement scheme. In Sec.\ \ref{IV}, we discuss the
measurement scheme written as one operator. In Sec.\ \ref{V}, we present a 
superposition of NOON states with infinite QFI and finite average photon 
number. Superpositions of NOON states in MMZI and dual Fock states in 
MZI are compared in Sec.\ \ref{VI}. Finally, a conclusion is given in
Sec.\ \ref{VII}.

\section{Simultaneous measurements}\label{II}
Two-path optical interferometers are widely studied to
estimate a completely unknown optical phase $\phi$ from the photon
numbers observed in two ports of the output.
Here we express the quantum mechanics of two-path interferometers
in terms of the spin-$N/2$ algebra of the Schwinger representation,
as shown in Ref.\ \cite{BCandGJ},
\begin{eqnarray}
\begin{array}{c}
\hat{J_{1}}=(\hat{a}^{\dag}\hat{b}+\hat{b}^{\dag}\hat{a})/2,\
\hat{J_{2}}=(\hat{a}^{\dag}\hat{b}-\hat{b}^{\dag}\hat{a})/2i,\\
\hat{J_{3}}=(\hat{a}^{\dag}\hat{a}-\hat{b}^{\dag}\hat{b})/2,
\label{spin-N/2 algebra}
\end{array}
\end{eqnarray}
where $\hat{a}$ and $\hat{b}$ are the annihilation operators of the
two paths. Eigenstates of $\hat {J}_{3}$ can be defined
as usual by Fock space representation as,
\begin{eqnarray}
|j,m\rangle=|j+m\rangle_{a}|j-m\rangle_{b}.
\end{eqnarray}
The phase shift of $\phi$ between the arms of the interferometer can
be expressed by a unitary transformation $\hat{U}_{\phi}=\exp(-i\phi\hat{J}_{3})$.
The unitary transformation of a beam-splitter is given by
$\hat{B}=\exp(i\pi\hat{J}_{1}/2)$.

\begin{figure}[!hbt]
\includegraphics[width=0.38\textwidth]{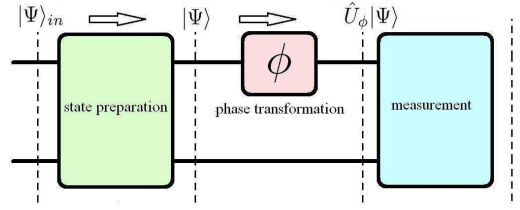}
\caption{(color online). General phase estimation procedure involving states preparation,
phase transformation and measurement.
\label{procedure}}
\end{figure}

Then, we consider that a linear phase estimation scheme in two-path
interferometers can be divided to three parts: states preparation,
phase transformation and measurement \cite{ouZY,LloydPRL,GOHLQM},
see Fig.\ \ref{procedure}. The procedure in Ref.\ \cite{nphi}
is a special example in which phase transformation repeats $n$
times during each measurement.

In this section, our discussion is focused on the measurement part.
We perform photon number counting measurements at two ports of linear
two-path interferometers to measure the output state $|\Psi\rangle_{out}$,
respectively, written as $\hat{n}_{a}$ and $\hat{n}_{b}$. It is obvious
that $\hat{n}_{a}$ and $\hat{n}_{b}$ are compatible observables,
$[\hat{n}_{a},\hat{n}_{b}]=0$. Thus, measurements $\hat{n}_{a}$ and $\hat{n}_{b}$
can be performed simultaneously. This pair of measurements can not be expressed
as $\hat{n}_{a}\otimes\hat{n}_{b}$ because we obtain two independent results
that can be written as a column matrix, $(n_{a},n_{b})^{T}$, from simultaneous
measurements of any two-mode Fock state,
\begin{eqnarray}
\left(
\begin{array}{c}
\hat{n}_{a}\\
\hat{n}_{b}
\end{array}\right)\sum_{j}\sum_{m=-j}^{j}C_{j,m}|j+m\rangle_{a}|j-m\rangle_{b}=\nonumber\\
\sum_{j}\sum_{m=-j}^{m=j}C_{j,m}
\left(
\begin{array}{c}
j+m\\
j-m
\end{array}\right)|j+m\rangle_{a}|j-m\rangle_{b},
\end{eqnarray}
where $\sum_{j}\sum_{m=-j}^{j}|C_{j,m}|^{2}=1$.
However, the operator $\hat{n}_{a}\otimes\hat{n}_{b}$ only obtains a result
$n_{a}\times n_{b}$, and is useless to describe a pair of simultaneous measurements.
Here we define simultaneous measurements as using several compatible observables
to perform a united measurement for a state at the same time. In fact, it is
sufficient and necessary for them to be measurable consecutively, such that the
joint probability of the results does not depend on the order of the measurement
\cite{book}.
Moreover, we show these pairs of simultaneous measurements are equivalent,
\begin{eqnarray}
\left(
\begin{array}{c}
\hat{n}_{a}\\
\hat{n}_{b}
\end{array}\right)\Leftrightarrow
\left(
\begin{array}{c}
\hat{n}_{b}\\
\hat{n}_{a}
\end{array}\right)\Leftrightarrow
\left(
\begin{array}{c}
\hat{N}\\
\hat{\Delta}
\end{array}\right)\Leftrightarrow
\left(
\begin{array}{c}
\hat{\Delta}\\
\hat{N}
\end{array}\right)
,
\end{eqnarray}
where $\hat{N}=\hat{n}_{a}+\hat{n}_{b}$ is total output photon
number, and $\hat{\Delta}=\hat{n}_{b}-\hat{n}_{a}$ is the difference
between the two ports.

Then, we show that the simultaneous measurements in two-path
interferometers can be described by two equivalent POVMs,
\begin{eqnarray}
\begin{array}{c}
\hat{E}(N,\Delta)\Leftrightarrow\hat{E}'(n_{a},n_{b})\\
\left|\frac{N}{2},\frac{\Delta}{2}\right\rangle\left\langle\frac{N}{2},\frac{\Delta}{2}\right|
\Leftrightarrow\left|\frac{N-\Delta}{2}\right\rangle_{a}\left|\frac{N+\Delta}{2}\right\rangle_{b}\left\langle\frac{N+\Delta}{2}\right|_{b}\left\langle\frac{N-\Delta}{2}\right|_{a}.
\end{array}
\label{operator}
\end{eqnarray}
$\{\hat{E}(n_{a},n_{b})\}$ and $\{\hat{E}(N,\Delta)\}$ are complete in the sense that,
\begin{eqnarray}
\sum_{n_{a},n_{b}}\hat{E}(n_{a},n_{b})=\sum_{N}\sum_{\Delta=-N}^{N}\hat{E}(N,\Delta)=\hat{I},
\end{eqnarray}
where $\hat{I}$ is a unit operator.
In subsequent sections, we use the equivalent simultaneous
measurements $\hat{E}(N,\Delta)$ for simpleness.

\section{Fisher information and Cram\'{e}r-Rao bound}\label{III}
To estimate the parameter $\phi$, a POVM $\{\hat{E}(\bm{\xi})\}$ is
performed on the output state, and $\phi$ is inferred from the measurement
results $\bm{\xi}$. The final measurement results are related
with Fisher information (FI) which takes the form,
\begin{eqnarray}
F_{\hat{E}(\bm{\xi})}[|\Psi_{\phi}\rangle]=\sum_{\bm{\xi}}P(\bm{\xi}|\phi)\left(\frac{\partial\ln{P}(\bm{\xi}|\phi)}{\partial\phi}\right)^{2},
\end{eqnarray}
where $|\Psi_{\phi}\rangle=\hat{U}_{\phi}|\Psi\rangle$, and likelihood function
$P(\bm{\xi}|\phi)=\langle\Psi_{\phi}|\hat{E}(\bm{\xi})|\Psi_{\phi}\rangle$
is the probability of measurement results $\bm{\xi}$ given that
the true system phase is $\phi$. CRB places a limit on
the mean of the square of the unbiased phase error via the Cram\'{e}r-Rao
inequality,
\begin{eqnarray}
\Delta\phi\geq1/\sqrt{F_{\hat{E}(\bm{\xi})}[|\Psi_{\phi}\rangle]}.
\end{eqnarray}
We note that FI is additive, and for $M$ repeated trials of
the same measurement, the FI is
$M\times{F}_{\hat{E}(\bm{\xi})}[|\Psi_{\phi}\rangle]$, which leads
to CRB on the phase uncertainty, $\Delta\phi\geq1/(
M\times{F}_{\hat{E}(\bm{\xi})}[|\Psi_{\phi}\rangle]) ^{1/2}$. QFI for state $|\Psi_{\phi}\rangle$ is the maximum,
\begin{eqnarray}
F_{Q}[|\Psi_{\phi}\rangle]:=\max_{\{ \hat{E}(\bm{\xi})\}
}{F}_{\hat{E}(\bm{\xi})}[|\Psi_{\phi}\rangle],
\end{eqnarray}
which is saturated by a particular POVM \cite{Caves2,same}.
In the following discussion, we show that the measurement with form
$\hat{E}(N,\Delta)$ in Eq.\ (\ref{operator}) saturates the QFI when state
$|\Psi\rangle$ is a superposed NOON state.

When total photon number
fluctuates, we consider a pure state generated from the entangled photon
source in Fig.\ \ref{f1b},
\begin{eqnarray}
|\Psi\rangle=\sum_{N}\sqrt{P(N)}|\psi_{N}\rangle,
\label{intermediate}
\end{eqnarray}
where $P(N)$ refers to the probability distribution of total photon number
$N$, and
$|\psi_{N}\rangle=\sum_{m=-{N}/{2}}^{{N}/{2}}\sqrt{p(m)}|{N}/{2},m\rangle$.
When the phase is $\phi$, we write the probability of
any output result $(N,\Delta)^{T}$ as
$P(N,\Delta|\phi)={_{out}}\langle\Psi|\hat{E}(N,\Delta)|\Psi\rangle_{out}$,
in which $|\Psi\rangle_{out}=\hat{B}\hat{U}_{\phi}|\Psi\rangle$ is the
output state.
For the output result $N=n_{a}+n_{b}$, the probability is
$P(N|\phi)=\sum_{\Delta=-N}^{N}{_{out}}\langle\Psi|\hat{E}(N,\Delta)|\Psi\rangle_{out}$.
The conditional probability is $P(\Delta|N,\phi)=P(N,\Delta|\phi)/P(N|\phi)$.
Because $[\hat{B},\hat{N}]=[\hat{U}_{\phi},\hat{N}]=0$, the
probability distribution of photon number does not change
after beam-splitter or phase, $P(N|\phi)=P(N)$ and
$\partial{P}(N|\phi)/\partial{\phi}=0$. Hence, FI can be written as,
\begin{eqnarray}
F_{\hat{E}(N,\Delta)}[|\Psi\rangle]=\sum_{N}P(N)F_{\hat{E}(\Delta)}[|\psi_{N}\rangle].
\end{eqnarray}
This means that the FI of the state with a fluctuating photon number
equals to the probability summation of each FI of the superposed
state which has a fixed photon number. As the situation that the
photon number is fixed, FI of $|\psi_{N}\rangle$ is
$F_{Q}[|\psi_{N}\rangle]={N}^{2}$ if and only if $|\psi_{N}\rangle$
is a NOON state, and FI of any other state is less than $N^{2}$.
Therefore, we have,
\begin{eqnarray}
F_{\hat{E}(N,\Delta)}[|\Psi\rangle]\leq\max_{|\Psi\rangle}F_{Q}[|\Psi\rangle]=\langle\hat{N}^{2}\rangle,
\label{MQFI}
\end{eqnarray}
where the equality can only be achieved when $|\Psi\rangle$ is a
superposition of NOON states. Moreover this is QFI, and
$\{\hat{E}(N,\Delta)\}$ is the optimal POVM. According to the
Cram\'{e}r-Rao inequality, we obtain CRB as
\begin{eqnarray}
\Delta\phi\geq1/\langle\hat{N}^{2}\rangle^{1/2}.
\end{eqnarray}
Therefore, we find the optimal measurement and optimal states for
the maximum QFI in linear two-path interferometers. We remark that
we can calculate FI by using either intermediate state
$\hat{U}_{\phi}|\Psi\rangle$ or final output $|\Psi\rangle_{out}$
in Fig.\ \ref{f1}. It makes no difference for the result when POVM
$\{\hat{E}({\bm{\xi}})\}$ is optimal, as pointed out in Ref.\ \cite{LloydPRL}.

Here we emphasize that CRB is asymptotically tight for unbiased estimation
with infinite trials of measurements, and it does not provide a rigorous
basis for HL when considering the prior probability distribution of
overall phase shift. Even so, QFI is still an extremely important quantity
in quantum physics. When the system interacts with the environment,
such as quantum noise and photon losses, QFI changes and the precision
of phase sensitivity is affected. It is shown that QFI flow, written
as $\partial_{t}F_{Q}[|\Psi\rangle]$, directly characterizes the
non-Markovianity of the quantum dynamics of open systems \cite{scp}.
Therefore, QFI can also be evaluated by the phase sensitivity in the
experiment to measure the non-Markovianity of quantum open systems.
Moreover, an environment-assisted precision measurement is also
available \cite{EAssisted}.

\section{Measurement as one observable estimator}\label{IV}
If a phase shift $\phi$ is measured by the outcomes of an observable
estimator $\hat{A}$ in linear two-path interferometers, the estimator
should be expressed as a binary function of $\hat{n}_{a}$ and
$\hat{n}_{b}$,
\begin{eqnarray}
\hat{A}=f(\hat{n}_{a},\hat{n}_{b}),
\end{eqnarray}
which is the case considered in Refs.\ \cite{Hofmann,novelty1,novelty2}.
Because simultaneous measurements are complete for the two-mode Fock
space, and are sufficient for phase $\phi$, we obtain,
\begin{eqnarray}
F_{\hat{A}}[|\Psi_{\phi}\rangle]\leq F_{\hat{E}'(\hat{n}_{a},\hat{n}_{b})}[|\Psi_{\phi}\rangle],
\label{inequal}
\end{eqnarray}
where the equal holds if the measurement $\hat{A}$ of the state
$|\Psi_{\phi}\rangle$ is a sufficient statistic for underlying parameter
$\phi$ \cite{ss}; for example, an observable $\hat{A}=\hat{n}_{a}+\sqrt{2}\hat{n}_{b}$ is sufficient for QFI for any
state in the two-mode Fock space.

Note that one observable estimator that is sufficient for QFI when
considering a fixed photon number may be insufficient when the photon
number is fluctuating. For example, 
if the total photon number is fixed and known to be $N$,
momentum operator $\hat{J}_{3}$ \cite{mini-uncertainty} is sufficient 
to achieve the maximum QFI when state
$|\Psi\rangle$ generated from the entangled photon source in MMZI
is a NOON state; but we confirm that
$\hat{J}_{3}$ is not the optimal measurement for QFI when
total photon number fluctuates. 

We should also notice that some operators, for example the parity 
operator $\hat{\Pi}_{a}=\exp{(i\pi\hat{n}_{a})}$ \cite{Gerry,Dowling}, 
are still useful in the quantum metrology for high precise
parameter estimation, although they can not saturate QFI for states 
with a fluctuating photon number. Moreover, when deriving the 
appropriate form of HL, it is reasonable to consider the situation 
that the parameter is estimated from the results of quantum measurement 
which is a sufficient statistic for the parameter.

\section{A superposition of NOON states with arbitrary high phase sensitivity and finite average photon number}\label{V}
\begin{figure}
\centering
\subfigure[]{
\includegraphics[width=0.35\textwidth]{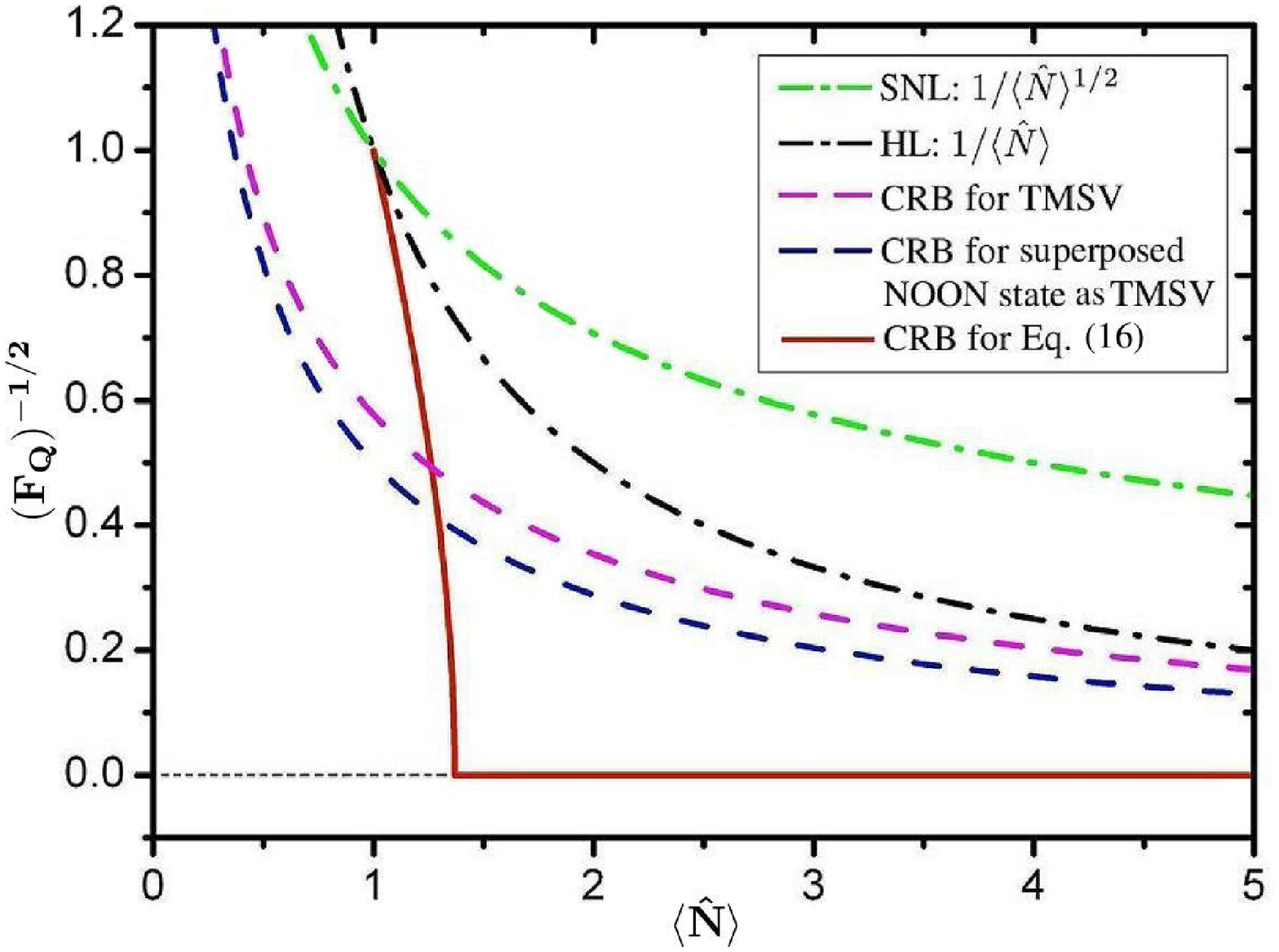}\label{f3a}}
\subfigure[]{
\includegraphics[width=0.35\textwidth]{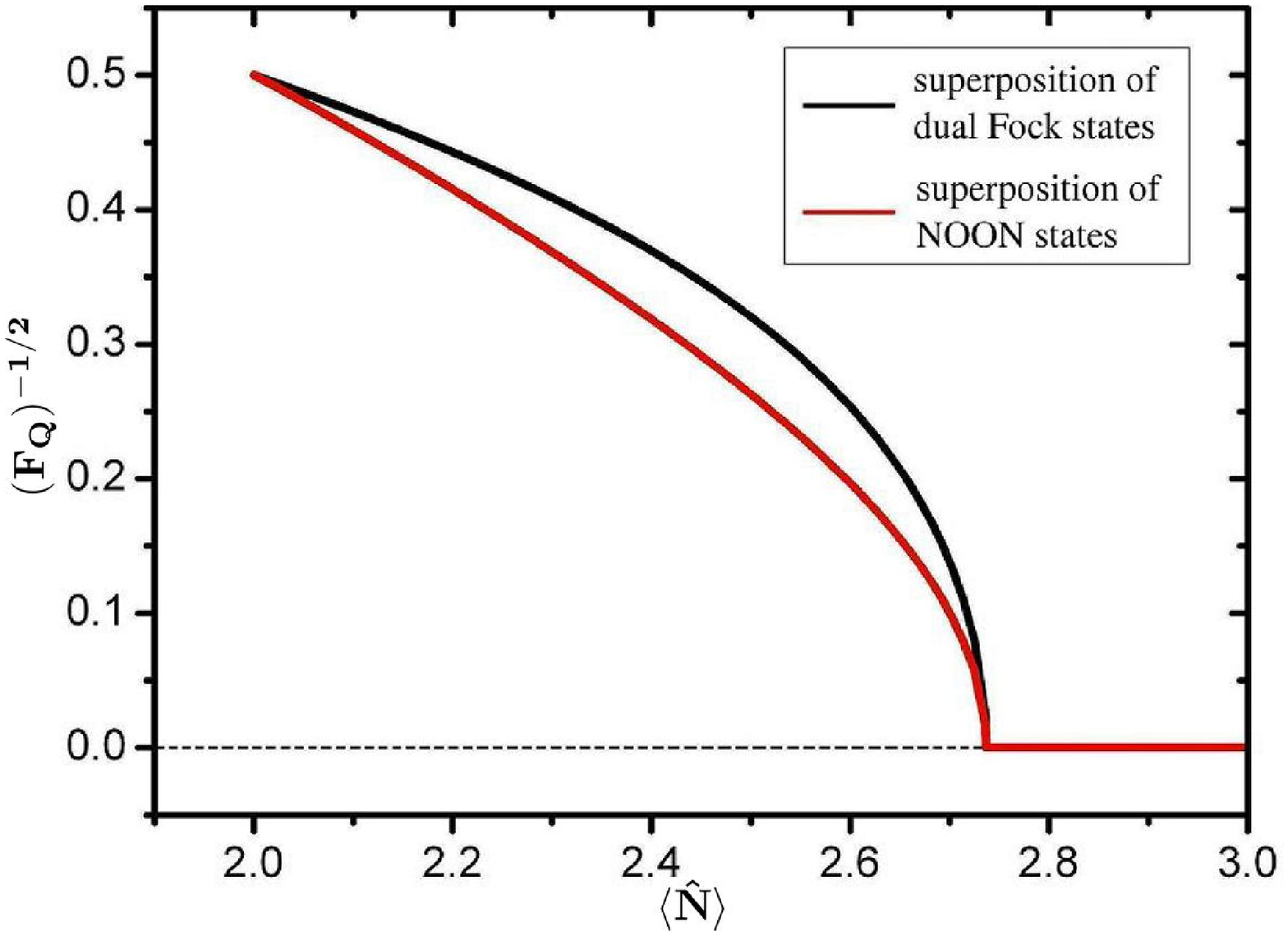}\label{f3b}}
\caption{(color online). (a)\ Phase uncertainty $\Delta\phi$ against
average photon number $\langle\hat{N}\rangle$. Green dash-dot line
is for the SNL $1/\langle\hat{N}\rangle^{1/2}$, black
dash-dot line is for conventional HL
$1/\langle\hat{N}\rangle$, purple
dash-dot line denotes CRB of TMSV dual Fock state $1/(\langle
\hat{N}^{2}\rangle +2\langle\hat{N}\rangle)^{1/2}$, blue dash line
indicates the minimum CRB in TMSV states $1/(2\langle
\hat{N}^{2}\rangle +2\langle\hat{N}\rangle)^{1/2}$, and red solid
line is for CRB of superposed NOON states in
Eq.\ (\ref{superNOON}).
When $\langle\hat{N}\rangle\geq1.369$, state in
Eq.\ (\ref{superNOON}) obtains an infinite QFI and a zero CRB.
(b)\ Phase uncertainty $\Delta\phi$ against average
photon number $\langle\hat{N}\rangle$. Red solid line is for the
CRB obtained by the superposition of NOON states
similar as Eq.\ (\ref{superNOON}), and black solid line is for the
superposition of dual Fock states. When
$\langle\hat{N}\rangle<2.737$, QFI for NOON state is
larger than dual Fock state, while when
$\langle\hat{N}\rangle\geq2.737$, both QFIs are
infinite.
\label{fig3}}
\end{figure}

As we have already shown, a superposition of NOON states will
always saturate the maximum QFI in Eq.\ (\ref{MQFI}). Here we present
an interesting example of such a state which has an infinite
QFI when the average photon number is finite. We consider a
superposition of NOON states as the following,
\begin{eqnarray}
|\Psi(x)\rangle=\frac{1}{\sqrt{\zeta(x)}}
\sum_{N=1}^{\infty}\frac{1}{\sqrt{N^{x}}}\frac{|N\rangle_{a}
|0\rangle_{b}+|0\rangle_{a}|N\rangle_{b}}{\sqrt{2}},
\label{superNOON}
\end{eqnarray}
where $x\in (1,+\infty)$,
$\zeta(x)=\sum_{N=1}^{\infty}1/N^{x}$ is Riemann Zeta
function,
and here it is the normalization factor. The
probability of each NOON state, $(|N\rangle_{a}
|0\rangle_{b}+|0\rangle_{a}|N\rangle_{b})/{\sqrt{2}}$, is $1/(\zeta(x)\times N^x)$
for photon number $N$. When $x\leq2$ the average of photon number
is infinite. We next consider the case $x=3$,
the photon number on average can be calculated as,
\begin{eqnarray}
\langle\hat{N}\rangle=\frac
{1}{\zeta(3)}\sum_{N=1}^{\infty}\frac{N}{N^{3}}=\frac{\zeta(2)}{\zeta(3)}\approx1.369.
\label{Naverage}
\end{eqnarray}
This means that a superposed state, $|\Psi(3)\rangle $ in Eq.\ (\ref{superNOON}),
has a finite average photon number, while the average of squared photon
numbers, $\langle\hat{N}^{2}\rangle$, can be infinite, as calculated
in the following,
\begin{eqnarray}
\langle\hat{N}^{2}\rangle=\frac
{1}{\zeta(3)}\sum_{N=1}^{\infty}\frac{N^2}{N^{3}}=\frac{\zeta(1)}{\zeta(3)}\rightarrow\infty,
\end{eqnarray}
where we have used the fact, $\zeta(1)\rightarrow \infty $.
For the general case, both $\langle\hat{N}\rangle$ and
$\langle{\hat{N}}^{2}\rangle^{1/2}$ are dependent on $x$, the
relation between the photon number on average
and CRB of phase estimation for state Eq.\ (\ref{superNOON}),
$\Delta\phi= 1/\langle\hat{N}^2\rangle ^{1/2}$, is shown in Fig.\ \ref{fig3}\ (a).
We can see from Fig.\ \ref{f3a} that CRB reaches zero when
$\langle\hat{N}\rangle$ approaches about 1.369, and as
$\langle\hat{N}\rangle$ increases, CRB remains as zero.

Note that in Refs.\ \cite{80s,ouZY}, a following state was proposed,
\begin{eqnarray}
|\Phi\rangle_{ssw}=A\sum_{m=0}^{M}\frac{1}{m+1}|m\rangle
,(M\gg1,A\simeq\sqrt{6/\pi^{2}}).
\end{eqnarray}
It is shown that QFI of this state can be arbitrary high
when the average photon number is infinite, while, in our case as
Eq.\ (\ref{superNOON}), the average photon number can be finite.

\section{Superpositions of NOON states and dual Fock states}\label{VI}
Dual Fock state, $|N\rangle_{a}|N\rangle_{b}$, is closely related with
NOON state and is also widely used in the optical interferometric
quantum measurement. For example, when input is a dual Fock state
$|1\rangle_{a} |1\rangle_{b}$ in MZI, the intermediate state is a
NOON state with photon number $N=2$, see \cite{USTC}. We next
compare QFI of superposition of dual Fock states
with that of NOON states in two examples.
We emphasize that NOON states and dual Fock
states are compared in different interferometers. NOON states are
generated by the entangled photon source in Fig.\ \ref{f1b},
while dual Fock states are the input states of MZI in Fig.\ \ref{f1a}.
The reason why we consider dual Fock states is that generating dual Fock
states is easier than generating the NOON states when photon number grows high.

The first example: we consider the input state as the two-mode
squeezed vacuum (TMSV), which is a superposition of dual Fock
states,
$|\Psi\rangle_{in}=\sum_{N=0}^{\infty}\sqrt{p_{N}(\langle\hat{N}\rangle)}|N\rangle_{a}|N\rangle_{b}$,
where
$p_{N}(\langle\hat{N}\rangle)=(1-t_{\langle\hat{N}\rangle})t_{\langle\hat{N}\rangle}^{N}$
with $t_{\langle\hat{N}\rangle}=1/(1+2/\langle\hat{N}\rangle)$, see
\cite{Dowling}. When a dual Fock state
$|\Psi\rangle_{in}=|N\rangle_{a}|N\rangle_{b}$ passes through the
first beam-splitter, the intermediate state is,
\begin{eqnarray}
&&|\Psi\rangle=\sum_{k=0}^{N}\frac{C_{N}^{k}\sqrt{(2N-2k)!(2k)!}}{N!2^{N+1/2}}
\times \nonumber \\
&&~~[(-1)^{N}|2N-2k\rangle_{a}|2k\rangle_{b}+|2k\rangle_{a}|2N-2k\rangle_{b}].
\label{dualfock}
\end{eqnarray}
QFI of each superposed state is $(2N-4k)^{2}$. Hence, QFI of the
intermediate state Eq.\ (\ref{dualfock}) can be calculated to be
$2N^{2}+2N$. Then QFI of the TMSV is
$\sum_{N=0}^{\infty}p_{N}(\langle\hat{N}\rangle)(2N^{2}+2N)=\langle\hat{N}\rangle^{2}+2\langle\hat{N}\rangle$.
In comparison, for a superposition of NOON states with a same
probability distribution as TMSV,
$|\Psi\rangle=\sum_{N=0}^{\infty}\sqrt{p_{N}(\langle\hat{N}\rangle)}(|2N\rangle_{a}|0\rangle_{b}
+|0\rangle_{a}|2N\rangle_{b})/\sqrt{2}$, the QFI is
$\sum_{N=0}^{\infty}p_{N}(\langle\hat{N}\rangle)4N^{2}
=2\langle\hat{N}\rangle^{2}+2\langle\hat{N}\rangle$. This state
saturates the maximum QFI for this probability distribution of total
photon number. CRBs of these two states are shown in
Fig.\ \ref{f3a}.

Second example: similar as the state in Eq.\ (\ref{superNOON}), we
consider a superposition of dual Fock states with the same
probability distribution of the photon number, $|\Psi(x)\rangle _{dualFock}=
\sum_{N=1}^{\infty}|N\rangle_{a}|N\rangle_{b}/\sqrt{\zeta(x)\times N^{x}}$.
When $x=3$, the average photon number is
$\langle\hat{N}\rangle=\frac{2\times{\zeta}(2)}{\zeta(3)}\approx2.737$,
and QFI can be calculated as,
$F_{Q}(\phi)=\sum_{N=1}^{\infty}\left[\frac{2\times{\zeta}(1)}{\zeta(3)}+\frac{2\times{\zeta}(2)}{\zeta(3)}\right]\rightarrow\infty$.
Subsequently, it is with a zero CRB similar as the state in
Eq.\ (\ref{superNOON}). The results of these two states
are plotted in Fig.\ \ref{f3b}. Note that for probability distribution
$1/(\zeta(x)\times N^x)$, the photon number of NOON state is $2N$ in
comparing with that of dual Fock state. We can see that when
$\langle\hat{N}\rangle<2.737$, QFI for NOON state case is larger than
case of dual Fock state; when $\langle\hat{N}\rangle\geq2.737$,
CRBs of both states are zeroes.

Generally, for any pure intermediate state written as Eq.\ (\ref{intermediate}),
the form of superposed state $|\psi_{N}\rangle$ and the distribution
of photon number $P(N)$ are both important to achieve a high QFI
in linear two-path interferometers, which can be regarded as the
effects of entanglement as well as superposition. A similar discussion
can also be found in Ref.\ \cite{cite}. Another interesting result is
that the states with infinite QFI might not have any Zeno dynamics
since the quantum Zeno time scale expressed in terms of QFI, written as $\tau_{\textrm{QZ}}=2/\sqrt{mF_{Q}}$ \cite{QZE}, approaches zero.

\section{Conclusion and Discussion}\label{VII}
We propose that the maximum QFI of states with a fluctuating or fixed
number of particles takes the form, $\langle\hat{N}^{2}\rangle$
in two-path interferometer. Our result shows the superposed NOON states
by simultaneous measurements achieve this optimal QFI in MMZI. We also
present a specified superposition of NOON states with probability
distribution related with Riemann Zeta function. This state has an
infinite QFI but with a finite average photon number via linear two-path
interferometer measurements. We also compare the case of dual Fock state
in MZI with that of NOON state in MMZI. Our work presents that the
advantage of this unbounded quantum Fisher information can be obtained
by only a few superposed terms, which will be beneficial to the future
development of quantum technology.

We assume that the entangled photon source in Fig.\ \ref{f1b}
could produce superposed NOON states. One area for future research
is to prepare superposed NOON states when some achievements have
been made for the formation of high-NOON states
\cite{generate1,generate2}.

\begin{acknowledgments}
We thank Lorenzo Maccone, Seth Lloyd and Michael Hall for useful discussions. The
authors are also grateful to Tian-Si Han and Si-Wen Li for discussions. We also 
acknowledge Referees for their useful suggestions which help us a lot to 
improve our work. This work is supported by NSFC (10934010, 10974233, 10974247), ``973''
program (2010CB922904, 2011CB921500).
\end{acknowledgments}

\end{document}